\documentclass[11pt]{article}

\usepackage{amsmath, amsthm, amsopn, amssymb, enumerate, ifpdf}
\usepackage[english]{babel}
\ifpdf
\usepackage[pdftex,urlcolor=blue]{hyperref}
\else
\usepackage[hypertex,urlcolor=blue]{hyperref}
\fi
\hypersetup{colorlinks}
\usepackage{geometry}
\newtheorem{thm}{Theorem}[]

\newtheorem{claim}[thm]{Claim}
\theoremstyle{definition}

\def\FullBox{\hbox{\vrule width 8pt height 8pt depth 0pt}}
\newcommand{\QED}{\;\;\;\FullBox}
\newenvironment{Proof}{\noindent{\bf Proof:~~}}{\hfill\QED}
\newcommand{\BPF}{\begin{Proof}} \newcommand {\EPF}{\end{Proof}}

\newcommand{\E}{\mathbb{E}}
\newcommand{\Var}{\text{Var}}
\renewcommand{\Pr}{{\rm Pr}}

\newcommand{\eqdef}{\stackrel{\rm def}{=}}

\DeclareMathOperator\Inf{Inf}
\renewcommand\emptyset{\phi}

\title{On the Converse of Talagrand's Influence Inequality}

\author{Saleet Klein\thanks{Tel Aviv University, \href{mailto:saleetklein@mail.tau.ac.il}{\nolinkurl{saleetklein@mail.tau.ac.il}}} \and Amit Levi\thanks{Tel Aviv University, \href{mailto:amitlev3@post.tau.ac.il}{\nolinkurl{amitlev3@post.tau.ac.il}}} \and Muli Safra\thanks{Tel Aviv University, \href{mailto:safra@cs.tau.ac.il}{\nolinkurl{safra@cs.tau.ac.il }}} \and Clara Shikhelman\thanks{Tel Aviv University, \href{mailto:clarashk@mail.tau.ac.il‬}{\nolinkurl{clarashk@mail.tau.ac.il‬}}} \and Yinon Spinka\thanks{Tel Aviv University, \href{mailto:yinonspi@post.tau.ac.il }{\nolinkurl{yinonspi@post.tau.ac.il}}} \\}

\date{}

\begin{document}
\maketitle
\begin{abstract}
In~\cite{Tal94}, Talagrand showed a generalization of the celebrated KKL theorem. In this work, we prove that the converse of this generalization also holds. Namely, for any sequence of numbers $0<a_1,a_2,\ldots,a_n\le 1$ such that $\sum_{j=1}^n a_j/(1-\log a_j)\ge C$ for some constant $C>0$, it is possible to find a roughly balanced Boolean function $f$ such that $\Inf_j[f] < a_j$ for every $1 \le j \le n$.
\end{abstract}
\section{Introduction}

In their seminal paper \cite{KKL88}, Kahn, Kalai and Linial showed that for any Boolean function $f$ there exists a coordinate $1 \le j \le n$ such that $\Inf_j[f]\ge c\cdot\frac{\log n}{n}\cdot \Var[f]$, where $c>0$ is a universal constant. This result, followed by the generalizations of Bourgain el al.\ \cite{BKK+92}, Talagrand \cite{Tal94} and Friedgut \cite{Fri98}, was a milestone for numerous results in different areas in computer science and mathematics such as hardness of approximation \cite{DS05,CKK+06,KR08}, distributed computing \cite{BL}, communication complexity \cite{Raz95}, metric embeddings \cite{KR09,DKSV06}, learning theory \cite{OS08,OW09}, random $k$-SAT \cite{Fri99}, random graphs \cite{FK96} and extremal combinatorics \cite{OW09}.

Talagrand's paper ``On Russo's approximate zero-one law" \cite{Tal94}, generalized KKL's result and stated that for every Boolean function $f$,
\[ \sum_{j=1}^n \frac{\Inf_j[f]}{1-\log\Inf_j[f]}\ge K\cdot \Var[f] ,\]
where $K>0$ is a universal constant.
We refer to this sum as \emph{Talagrand sum}.

We study whether the converse Talagrand's theorem holds. Namely, given a sequence of numbers, $0 < a_1,a_2,\ldots,a_n \le 1$ whose Talagrand sum is greater than a constant $C>0$, can one find a roughly balanced Boolean function $f$ such that $\Inf_j[f] < a_j$ for all $1 \le j \le n$. We show that this is true (up to a constant) not only for balanced functions, but also for unbalanced functions. 
\section{Main Result}
Let $n$ be a positive integer (which is henceforth fixed).
A {\em Boolean function} on $n$ variables is a function $f \colon \{0,1\}^n \to \{0,1\}$.
Let $x$ be a uniformly chosen element in $\{0,1\}^n$.
For $1 \le j \le n$, denote by $x^j$ the vector $x$ with the $j$-th coordinate flipped, i.e.,
\[ x^j \eqdef (x_1, \dots, x_{j-1}, 1-x_j, x_{j+1}, \dots, x_n) .\]
We define the {\em influence} of the $j$-th variable on $f$ to be
\[ \Inf_j[f] \eqdef \Pr[f(x) \neq f(x^j)] .\]
That is, $\Inf_j[f]$ is the probability that flipping the $j$-th bit affects the outcome of the function. 
We say that a Boolean function $f$ is {\em balanced} if $\Pr[f(x)=0]=1/2$. Throughout this paper, $\ln$ denotes the natural logarithm and $\log$ denotes the base-$2$ logarithm.

\begin{thm}
Let $0 < a_1,a_2,\ldots,a_n \le 1$ and denote
\[ \alpha \eqdef \sum_{j=1}^n\frac{a_j}{1-\log a_j} - 2 \ln 2 .\]
Then, for any $0<\mu \le 1-\exp(-\alpha/8)$, there exists a Boolean function $f$ on $n$ variables such that
\[ \mu \leq \E[f(x)] \le \frac{3}{4} \mu + \frac{1}{4} \]
and
\[ \Inf_j[f] < a_j \quad\text{ for all } 1 \le j \le n.\]
\end{thm}
\BPF
Following Ben-Or and Linial's example of tribes \cite{BL}, which minimizes the influence of each variable, we wish to construct a function in a similar manner by aggregating variables $x_j$ into tribes of various sizes according to the desired bound $a_j$ on their influence.

We may assume without loss of generality that $a_1 \ge a_2 \ge \cdots \ge a_n$. Define an integer $m \geq 0$ and a sequence of integers $k_0,k_1,\dots,k_m,k_{m+1}$ by
\begin{align*}
k_0 &\eqdef 0, \\
k_i &\eqdef \min \{ k \ge 1 ~:~ a_{k_0+\cdots+k_{i-1}+k} > 2^{1-k} \} , \quad 1 \leq i \leq m , \\
k_{m+1} &\eqdef n - (k_0 + \cdots + k_m) + 1 ,
\end{align*}
where $m$ is defined by the condition $\{ k \ge 1 ~:~ a_{k_0+\cdots+k_m+k} > 2^{1-k} \} = \emptyset$. Note that $2 \leq k_1 \le k_2 \le \cdots \le k_m$.
The integers $k_1,\dots,k_m$ represent the sizes of the tribes.
We shall show that there exists an integer $0 \leq m_* \leq m$ such that the function $f \colon \{0,1\}^n \to \{0,1\}$ defined by
\begin{equation}\label{eq:def-f}
f(x_1,\dots,x_n) \eqdef \bigvee_{i=1}^{m_*} \bigwedge_{j=1}^{k_i} x_{k_1+\cdots+k_{i-1}+j}
\end{equation}
satisfies the conclusion of the theorem.
We first show the following.
\begin{claim} \label{clm_expectation}
\[ \sum_{i=1}^{m} 2^{-k_i} \ge \frac{\alpha}{8} . \]
\end{claim}

\BPF
For $0 \leq i \leq m$, denote $s_i \eqdef k_0+\cdots+k_i$.
Note that, by the definition of $k_i$, we have
\[ a_{s_i + j} \le 2^{1-j} , \quad 0 \le i \le m,~ 1 \leq j \leq k_{i+1} - 1 \]
and, in particular,
\[ a_{s_i} \le a_{s_i - 1} = a_{s_{i-1}+k_i-1} \le 2^{2-k_i}, \quad 1 \le i \le m . \]
Therefore, since $x/(1- \log x)$ is increasing on $[0,1]$,
\begin{align*}
\sum_{j=1}^n\frac{a_j}{1-\log a_j}
 &= \sum_{j=1}^{k_1-1} \frac{a_j}{1-\log a_j}
  + \sum_{i=1}^{m} \sum_{j=0}^{k_{i+1}-1} \frac{a_{s_i+j}}{1-\log a_{s_i+j}} \\
 &\le \sum_{j=1}^{k_1-1} \frac{a_j}{1-\log a_j}
  + \sum_{i=1}^{m} \left( \sum_{j=0}^{k_i-1} \frac{a_{s_i}}{1-\log a_{s_i}} + \sum_{j=k_i}^{k_{i+1}-1} \frac{a_{s_i+j}}{1-\log a_{s_i+j}} \right) \\  
  &\le \sum_{j=1}^{k_1-1} \frac{2^{1-j}}{j}
  + \sum_{i=1}^{m} \left( k_i \frac{2^{2-k_i}}{k_i-1} + \sum_{j=k_i}^{k_{i+1}-1} \frac{2^{1-j}}{j} \right) \\  
  &= \sum_{j=1}^n \frac{2^{1-j}}{j}
  + 4 \sum_{i=1}^{m} \frac{k_i}{k_i-1} 2^{-k_i} \\
  &\le 2\ln 2 + 8 \sum_{i=1}^{m} 2^{-k_i} .
\end{align*}
Thus, the claim follows by the definition of $\alpha$.
\EPF

\medbreak

We construct the function as described in \eqref{eq:def-f}, where we keep adding tribes until we reach a point at which the expected value becomes at least $\mu$. We denote this tribe by $m_*+1$, so that at the end of the process we have $m_*$ tribes. Precisely, $m_*$ is defined by
\[ m_* \eqdef \min\left\{ 1 \le r \le m ~:~ \prod_{i=1}^r (1-2^{-k_i}) \leq 1-\mu \right\} .\]
Indeed, this is well-defined, since by Claim~\ref{clm_expectation},
\[ \prod_{i=1}^m (1-2^{-k_i}) \le \exp\left(- \sum_{i=1}^m 2^{-k_i} \right) \le \exp\left(-\frac{\alpha}{8}\right) \le 1-\mu .\]
Thus, with this choice of $m_*$, we have
\[ \E[f(x)] = \Pr[f(x)=1] = 1 - \Pr[f(x)=0] = 1 - \prod_{i=1}^{m_*} (1-2^{-k_i}) \ge \mu .\]
%
%
We now show that the expectation of the function $f$ constructed above is less than $\frac{3}{4} \mu + \frac{1}{4}$.
Indeed, by the minimality of $m_*$, and since $k_i \geq 2$ for all $1 \leq i \leq m$, we have
\[ \E[f(x)] = 1 - \prod_{i=1}^{m_*} (1-2^{-k_i}) < 1 - (1-\mu) (1-2^{-k_{m_*}}) \le 1 - \frac{3}{4}(1-\mu) = \frac{3}{4} \mu + \frac{1}{4} .\]

It remains to check that the influence of the $j$-th variable on $f$ is bounded above by $a_j$.
Indeed, if $s_{i-1} < j \leq s_i$ for some $1 \le i \le m_*$, then
\[ \Inf_j[f] = 2^{1-k_i} \prod_{\substack{\ell=1\\\ell \neq i}}^{m_*} (1-2^{-k_\ell}) \le 2^{1-k_i} < a_{s_i} \leq a_j .\]
Since $\Inf_j[f] = 0$ for all $s_{m_*} < j \le n$, the proof is complete.
\EPF

\section{Acknowledgments }
The authors would like to thank Gil Kalai for pointing out this question and to Guy Kindler for useful discussions.

\end{document}